\documentclass[prb,floatfix,showpacs,twocolumn,preprintnumbers,amsmath,amssymb,superscriptaddress,longbibliography,floatfix]{revtex4-2}
\usepackage{tabularx}
\usepackage{graphicx}
\usepackage{color}
\usepackage[utf8]{inputenc}
\usepackage{siunitx}

\usepackage{subcaption} % put this in your preamble

\usepackage{bm}
\usepackage{hyperref}
\usepackage{amsmath}
\usepackage{amsmath}
\let\oldAA\AA
\renewcommand{\AA}{\text{\normalfont\oldAA}}

\usepackage{orcidlink}

\usepackage{xcolor}
\colorlet{RED}{red}
\colorlet{BLUE}{blue}
\usepackage{booktabs, multirow}

\usepackage{xspace}
\newcommand*{\MS}{MoS$_2$\xspace}

\renewcommand{\AA}{\text{\normalfont\oldAA}}

\begin{document}

\newcommand{\figurewidth}{2.8in}

\title{Edge Magnetism in Colloidal \MS Triangular Nanoflakes}
\author{Surender Kumar\orcidlink{0009-0000-3072-5633}}
\email{surendermohinder@gmail.com}
\affiliation{Institut für Festk\"orpertheorie und -Optik, Friedrich-Schiller-Universit\"at Jena, 07743 Jena, Germany}

%%%%%Stefan
\author{Stefan Velja\orcidlink{0009-0003-1268-6273}}
\affiliation{Institut für Festk\"orpertheorie und -Optik, Friedrich-Schiller-Universit\"at Jena, 07743 Jena, Germany}

%%% Sufyan
\author{Muhammad Sufyan Ramzan\orcidlink{0000-0002-5017-8718}}
\affiliation{Institut für Festk\"orpertheorie und -Optik, Friedrich-Schiller-Universit\"at Jena, 07743 Jena, Germany}

\author{Caterina Cocchi\orcidlink{0000-0002-9243-9461}}
\email{caterina.cocchi@uni-jena.de}
\affiliation{Institut für Festk\"orpertheorie und -Optik, Friedrich-Schiller-Universit\"at Jena, 07743 Jena, Germany}

%%%%%%%% abstract %%%%%%
\begin{abstract}
The control of localized magnetic domains at the nanoscale holds great promise for next-generation spintronic applications. Colloidal transition metal dichalcogenides nanostructures are experimentally accessible and chemically tunable platforms for spintronics, deserving dedicated research to assess their potential. Here, we investigate from first principles free-standing triangular \MS nanoflakes with sulfur-terminated, hydrogen-passivated edges, to probe intrinsic spin behavior at varying side lengths. We find a critical edge length of approximately 1.5~nm separating nonmagnetic nanoflakes from larger ones with a magnetic ground state emerging from several, energetically competing spin configurations. In these systems, the magnetic activity is not uniformly distributed along the edges but localized on specific ``magnetic islands'' around molybdenum edge atoms. The localization of magnetic moments is robust even in non-equilateral nanoflake geometries, highlighting their intrinsic stability regardless of the (high) symmetry of the hosting structure. Our findings establish that the S-terminated, H-passivated triangular \MS nanoflakes are an energetically stable and potentially accessible platform via colloidal synthesis for low-dimensional, next-generation spintronic devices.

\end{abstract}

\maketitle

\section{Introduction}
The rich physics of spin-active materials has opened opportunities for spintronic applications, including high-density memory, logic circuits, and quantum information devices~\cite{Fabian2004,BHATTI2017,Ahn2020,Lin2019,Avsar2020}. With the emergence of edge-spin quantum magnetism in confined graphene nanostructures~\cite{Fern2007,Jiang2007,Sun2017a,Blackwell2021,Yazyev2010}, an increasing number of two-dimensional (2D) materials are explored for their potential use in spintronics~\cite{Liu2020,Brito2022,Tiutiunnyk2022,Abdel2023,SETHULAKSHMI2019}. Transition metal dichalcogenides (TMDs) are looked upon as the most promising candidates, thanks to their combined stability and structural flexibility even in confined nanostructured configurations~\cite{Liu2020,Brito2022,Tiutiunnyk2022,Abdel2023}. 

The magnetic behavior typical of materials with reduced dimensionality, has been often observed in TMD nanostructures~\cite{Li2008,Botello2009,Ataca2011,Deng2020,Shinde2019,Cui2017,Gibertini2015,Kou2012,Pan2012,Hui2012}. The local spin domains emerging in these systems is related to the presence of edges, which create a polar discontinuity~\cite{Gibertini2015,Gibertini2014}. 
The resulting electrostatic potential gradient drives the accumulation of charge carriers toward the edge to screen the polarization charges, ultimately leading to an edge-localized magnetization~\cite{Gibertini2015,Gibertini2014,Pino2020}.
As a result, while bulk TMDs are typically non- or very weakly magnetic~\cite{Rasmussen2015}, edge effects and lateral confinement can induce magnetism, with persistent ferromagnetic ordering even at room temperature~\cite{Dey2024,Li2008,Botello2009,Ouyang2014}. For example, TMD nanoribbons with armchair edges are typically nonmagnetic or weakly magnetic~\cite{Li2008,Botello2009,Ouyang2014}, while those featuring zigzag or sawtooth edges stabilize in a magnetic ground state~\cite{Li2008,Botello2009,Ataca2011,Deng2020,Shinde2019,Cui2017,Gibertini2015,Kou2012,Pan2012,Hui2012}. In these nanostructures with extended zigzag edges, the magnetic moment is generally delocalized along the entire edge, limiting the density of individually addressable magnetic elements~\cite{Li2008,Botello2009,Ataca2011,Deng2020,Shinde2019,Cui2017,Gibertini2015,Kou2012,Pan2012,Hui2012}. 
The lack of controllable, localized spin centers remains a critical challenge for utilizing TMDs in high-density spintronic devices~\cite{Feng2017,Ortiz2024}

Magnetism in TMDs can arise either via intentional introduction of defects~\cite{Fu2020,Guguchia2018,Liang2021,Zhang2024,Hossen2024,jaglivcic2003,mihailovic2003,gao2015,sanikop2019} or intrinsically from edge boundaries in confined nanostructures~\cite{Hui2012, Li2008,Botello2009,Ataca2011,Deng2020,Shinde2019,Cui2017,Gibertini2015,Kou2012,Pan2012}. Extrinsic magnetism, introduced by chalcogen vacancies~\cite{sun2014,mathew2012} or transition-metal dopants~\cite{Tian2015,martinez2018,Fu2020,wang2017}, generates localized spin domains. However, this approach offers limited control of the resulting magnetic properties, which exhibit a sensitive and often unpredictable dependence on the precise concentration, position, and clustering of the dopants. In contrast, intrinsic magnetism is inherently more robust and reliable as it is directly associated with the structural properties of the confined materials, including their edges~\cite{Fern2007,Jiang2007,Sun2017a,Blackwell2021,Yazyev2010}. For magnetic materials, a high density of reactive edges is distinctly advantageous. Edge atoms, due to changes in coordination, often deviate from the bulk-stoichiometric equilibrium, leading to atomic reconfigurations and non-uniform spin distributions~\cite{Fern2007,Jiang2007,Sun2017a,Blackwell2021,Yazyev2010}.

Triangular TMD nanoflakes can be highly promising spin hosts, 
as they exhibit intrinsic magnetism with localized magnetic moments~\cite{Tiutiunnyk2022,Abdel2023,cao2015,Liang2015,Pavlovi2015,Chen2018}, similar to their graphene-based analogs~\cite{Ortega2013,Maruyama2016,Han2024}. Colloidal triangular nanostructures offer the advantage to be synthesized with a uniform elemental termination and a single type of edge (e.g., zigzag) along all sides~\cite{Tiutiunnyk2022,Abdel2023,cao2015,Liang2015,Pavlovi2015,Chen2018}. While previous work on TMD nanoribbons focused on delocalized edge spins~\cite{Gibertini2014,Gibertini2015,Pino2020}, the confined geometry of triangular nanoflakes provides a unique structural platform to stabilize discrete, addressable magnetic units. This feature is crucial, as nanoflakes of other shapes (e.g., rectangular or rhombic) inherently result in a mixture of terminating elements and/or edge types. In TMDs, the structural characteristics of triangular quantum dots prevent the inherent limitations of nanoribbons or nanosheets, where opposite boundaries must contain a mixture of transition-metal ($d$-orbital magnetism) and chalcogen ($p$-orbital magnetism) terminations~\cite{Li2008, Castenetto2023,Ataca2011,Peter2019}, or where engineering uniform elemental termination forces the adoption of the armchair orientation on one side\cite{Li2008,Pan2012,Ridolfi2017}. By overcoming these constraints and ensuring uniform magnetic zigzag edges, triangular geometries circumvent constraints imposed by mixed terminations and nonmagnetic regions, making them a viable platform for generating stable, well-defined spin-active islands compared to conventional ribbon-like structures. 

In this work, we investigate the magnetic properties of colloidal \MS nanoflakes of triangular shape with hydrogen-passivated sulfur edges. While theoretical studies on similar systems based on first-principles and tight-binding models have been published in the last decade~\cite{Tiutiunnyk2022,Abdel2023,cao2015,Liang2015,Pavlovi2015,Chen2018}, we propose a systematic analysis to identify well-defined spin domains intrinsically arising in these systems due to their characteristic topology. By performing density functional theory (DFT) calculations, we show \MS nanoflakes with edge lengths smaller than 1.5~nm  are nonmagnetic, whereas larger structures exhibit energetically competing magnetic states separated by less than 50~meV, with varying total magnetic moments due to distinct local spin alignments. Despite S-termination and H-passivation, only specific Mo atoms close to the edges develop local magnetic moments forming symmetric topologies, in pronounced contrast to the delocalized spin density characterizing periodic TMD nanoribbons. Our results highlight triangular \MS nanoflakes with S-H edges as robust, experimentally feasible platforms to create high-density, individually addressable magnetic islands for spin-filtering and next-generation spintronic applications.

\begin{figure}
\centering
  \includegraphics[width=0.5\textwidth]{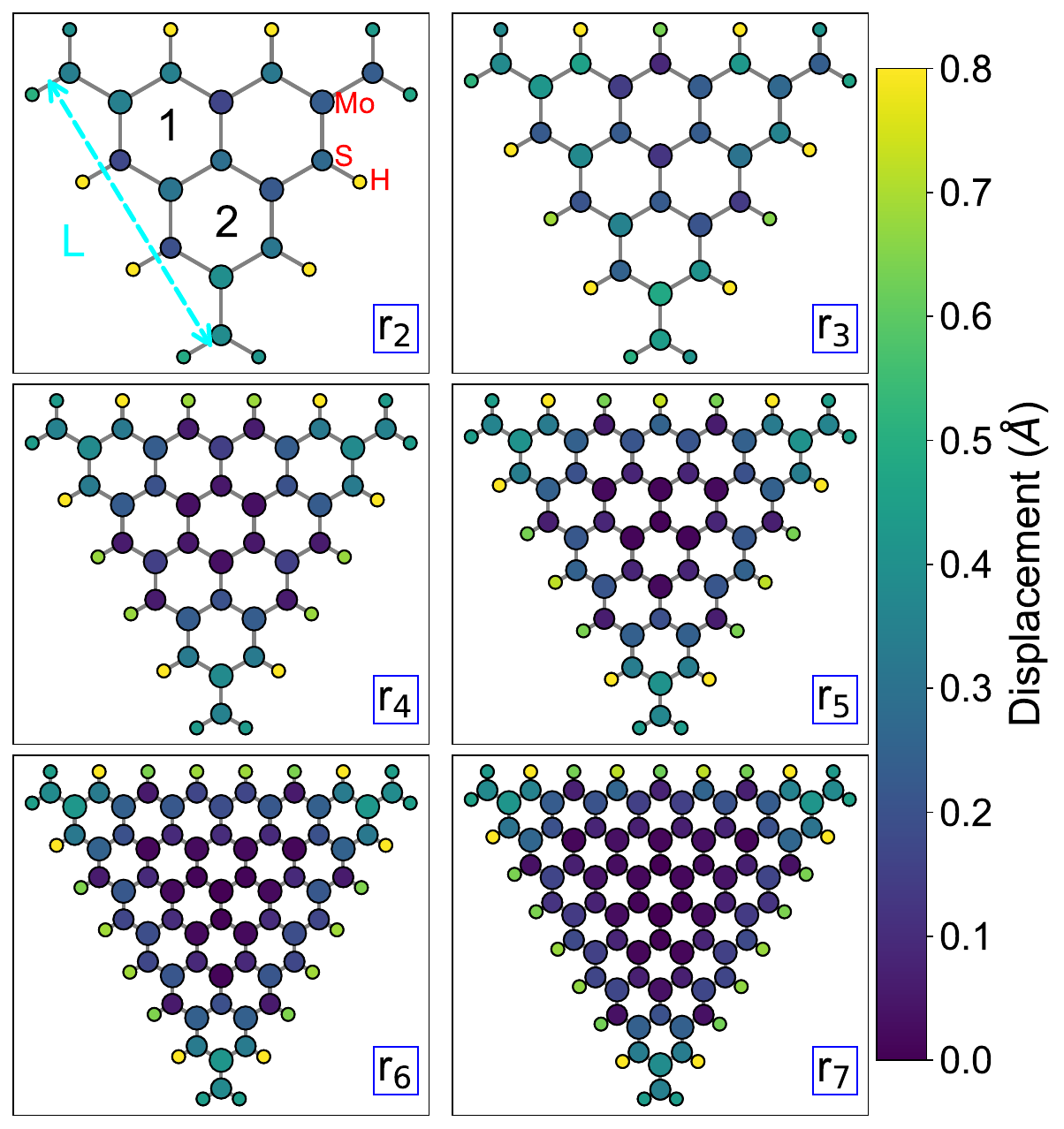}
\caption{Hydrogen-passivated triangular \MS nanoflakes, with an increasing number of rings along the edge length (L) and labeled according to them. The atoms are color-coded according to the magnitude of their displacement from the initial positions after structural optimization.} 
\label{fig:displaced_str}
\end{figure}
\section{Computational Details}
\label{sec:theory}
All calculations presented in this work were performed with spin-polarized (SP)  DFT~\cite{Hohenberg1964} as implemented in the Vienna Ab-initio Simulation Package (VASP)~\cite{Kresse1996} within the projector augmented wave framework~\cite{blochl1994,Kresse1999}. We used the generalized gradient approximation in the Perdew-Burke-Ernzerhof (PBE) implementation~\cite{Perdew1996}, incorporating Grimme’s D3 dispersion correction and Becke-Johnson damping~\cite{Grimme2010,Grimme2011}. Since the considered nanoflakes are non-periodic, we sample the Brillouin zone only at the  $\Gamma$-point, and add to the simulation box a vacuum layer of 16~\AA{} in all directions, to eliminate spurious interactions between periodic images of the unit cell. We relax the initial structures imposing SP until the Hellmann-Feynman forces acting on each atom were lower than 0.005~eV\AA$^{-1}$. The total energy convergence criterion for the self-consistent runs is set to 10$^{-8}$~eV and the plane-wave basis-set cutoff energy to 475 eV.

\begin{figure}%[b]
  \centering
  \includegraphics[width=0.5\textwidth]{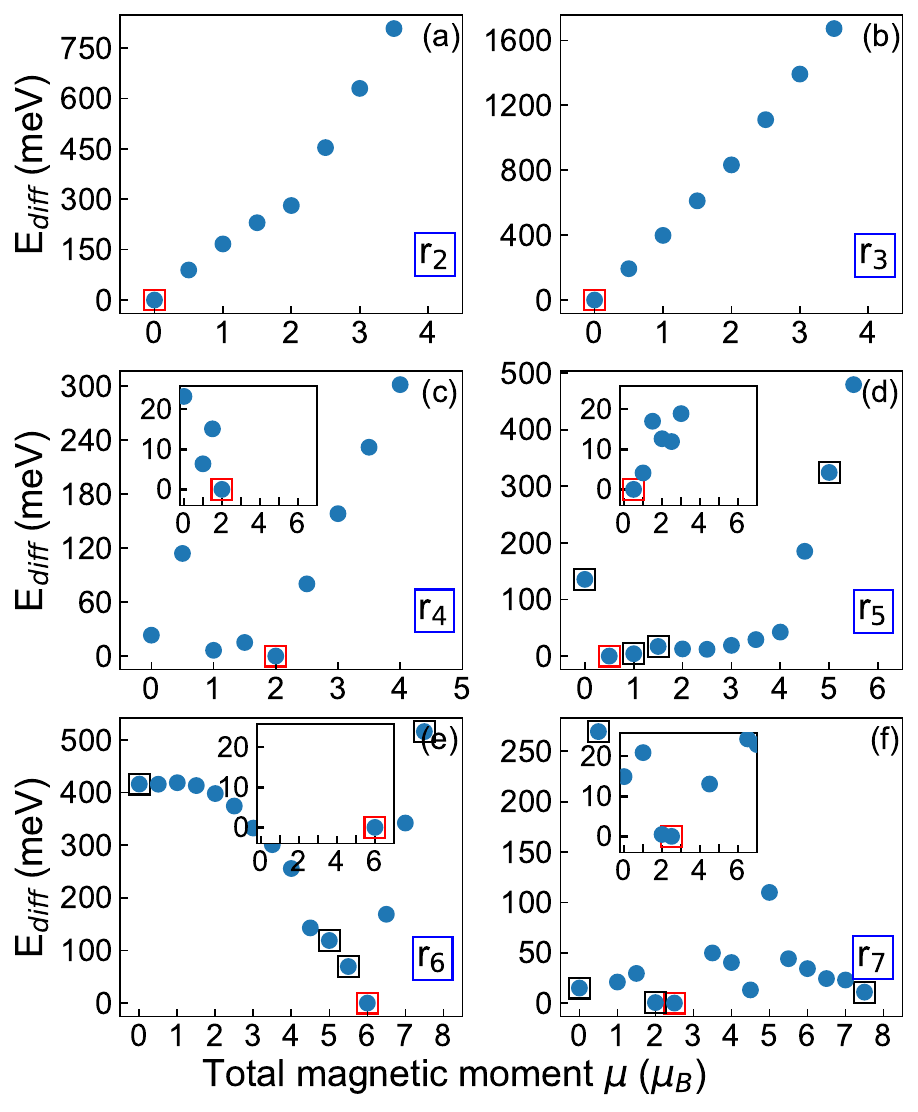}
  \caption{ Total energy difference ($E_{diff}$) relative to the magnetic ground state (red square) for various fixed total magnetic moments ($\mu$) of the considered \MS nanoflake. Less stable configurations discussed in the text are squared in black. Insets: Zoom-in to nearly degenerate magnetic orderings. }
  \label{fig:mag_mom_en}
\end{figure}

Due to the possible coexistence of multiple local minima predicted by SP-DFT~\cite{Huebsch2021}, the final magnetic state of a system depends on the starting spin configuration. To account for this uncertainty, we perform our calculations fixing the targeted total magnetic moments while allowing the system to self-consistently rearrange its local spins to satisfy the imposed constraint for the total magnetization $\mu$. This strategy minimizes the dependence on the initial conditions, enabling a systematic energetic comparison among different magnetic states to identify the ground state.

\section{Results and discussion}
\subsection{Structural properties}
In this work, we investigate six equilateral triangular \MS nanoflakes with varying edge lengths (L) generated from the 2H monolayer with experimental lattice parameters $a = 3.16$~\AA{}~\cite{Wildervanck1964, Coehoorn1987,Dungey1998}. The nanoflakes are labeled according to the number of \MS rings along their edge (Figure~\ref{fig:displaced_str}), ranging from the smallest nanoflake with two rings (r$_2$) to the largest one with seven rings (r$_7$). Their lateral sizes span approximately 15~\AA{}, ranging from 9.5~\AA{} for r$_2$ to 25.2~\AA{} for r$_7$. In each nanoflake, the interior molybdenum (Mo) atoms are sixfold coordinated by sulfur (S), whereas each interior sulfur atom is threefold coordinated by Mo. The edges are terminated by hydrogen-passivated sulfur atoms, which are known to be chemically stable in colloidal \MS nanostructures~\cite{Seyyedmajid2018,Lauritsen2003,Topsoe1993,Kyung2017,Din2021}.

We first briefly summarize the impact of structural relaxation by analyzing atomic displacements with respect to the initial positions across all nanoflakes (Figure~\ref{fig:displaced_str}). Note that all atoms were allowed to fully relax, and displacements of the final ($f$), optimized structures are calculated as the Euclidean distance $d = \sqrt{(x_f - x_i)^2 + (y_f - y_i)^2 + (z_f - z_i)^2}$ relative to the unrelaxed, initial ($i$) structure. As expected, edge atoms experience the largest shifts, ranging from 0.3--0.8~\AA{}, due to their modified coordination environments. Specifically, H atoms exhibit the most significant displacements (0.5--0.8~\AA{}), owing to their small size and arbitrary initial placement as passivating species. Both Mo and S atoms on the corners show moderate to large positional shifts, on the order of 0.5~\AA{}, whereas atoms in the core are only negligibly affected, with average displacements below 0.2~\AA{}. The atoms undergo highly symmetric displacements, as expected from the equilateral geometry of the structure. As the system size increases, a clear distinction emerges between bulk-like atoms, which exhibit small displacements, and edge-like atoms, which show significantly larger displacements (Figure~\ref{fig:displaced_str}).

The analysis of the bond distances (Figure~S1) reveals a narrow distribution for the Mo-S and S-H bonds, confirming the structural stability of both the core and the H-passivated edges. In contrast, the Mo-Mo distances vary slightly, within 10\% of the experimental value of (3.16 \AA{}) of the 2D \MS sheet~\cite{Wildervanck1964, Coehoorn1987,Dungey1998}), while S-S distances show the largest deviation, up to approximately 0.6~\AA{}, or around 20\% of the lattice parameter of \MS monolayer. This pronounced variation stems from the unique H-passivated environment of the S atoms at the nanoflake corners, which induces enhanced structural relaxation compared to the interior (Figure~\ref{fig:displaced_str}).
\begin{figure}
  \centering
  \includegraphics[width=.5\textwidth]{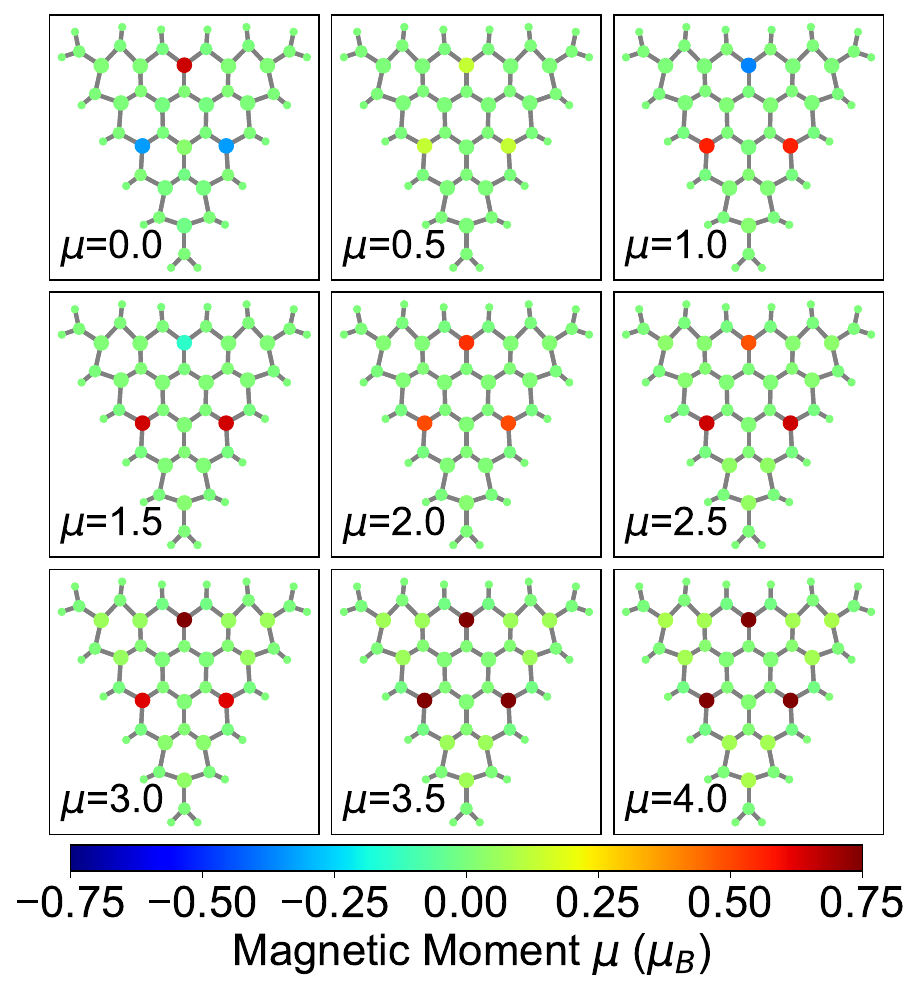}
    \caption{Spatial distribution of the local magnetic moments in the r$_4$ nanoflake at varying  total magnetization (in $\mu_B)$ indicated in each panel. The color scale for the atoms represents the magnitude and orientation of the local spins.}
  \label{fig:r4_mag_mom}
\end{figure}
\begin{figure*}
  \centering
  \includegraphics[width=0.95\textwidth]{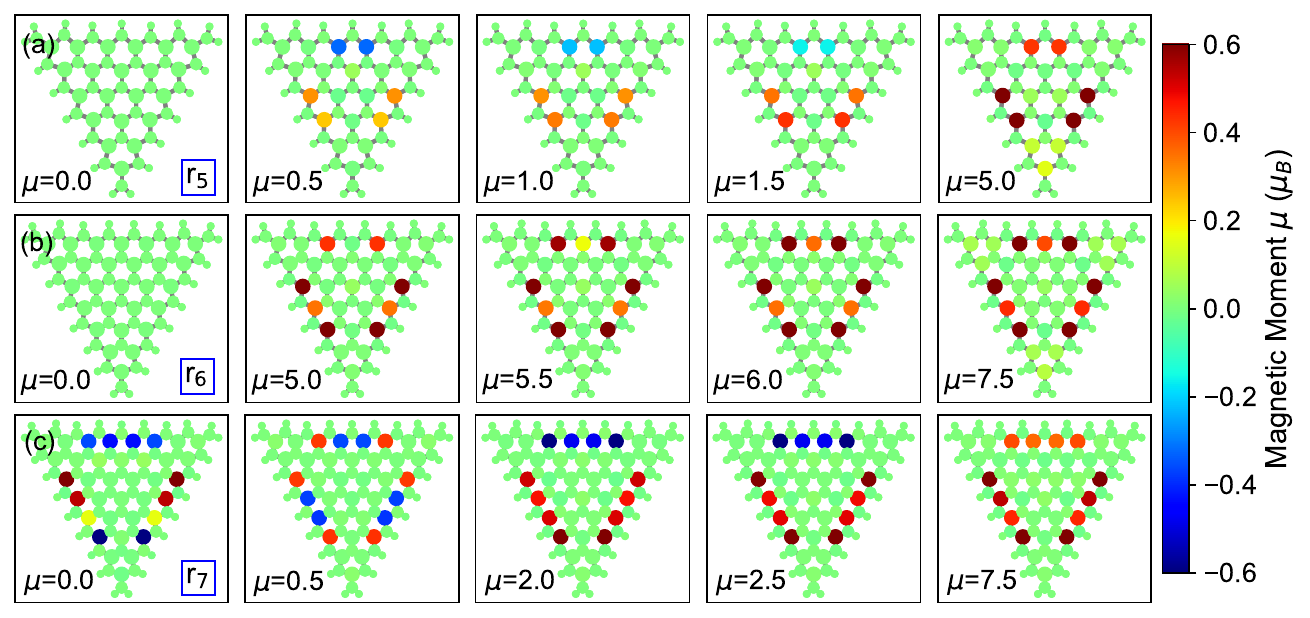}
  \caption{Spatial distribution of the local magnetic moments in selected (a) r$_5$, (b) r$_6$, and (c) r$_7$ magnetic configurations. The color scale for the atoms represents the magnitude and orientation of the local spins.}
  \label{fig:mag_selected}
\end{figure*}

\subsection{Magnetic ground-state determination}
We next investigate different magnetic configurations of the relaxed structure for each system size to determine their ground state (see Section~\ref{sec:theory} for details on the procedure). As shown in Figure~\ref{fig:mag_mom_en}, where the calculated total energy differences relative to the most stable magnetic state (red squares) are displayed, the smallest nanoflakes (r$_2$ and r$_3$) possess a non-magnetic ground state. In these systems, states with higher magnetic moments lead to a sharp increase in total energy, on the order of 800~meV and 1.6~eV for the largest explored magnetization, $\mu = 4 \; \mu_B$ (Figure~\ref{fig:mag_mom_en}, top panels), confirming the superior stability of the non-magnetic phase.
In contrast, beginning with r$_4$ and for all larger nanoflakes, magnetic configurations are energetically favored, suggesting 1.5~nm as a size threshold for the appearance of a magnetic ground state, similar to graphene nanoislands~\cite{Fern2007}. In the larger nanoflakes (r$_4$ to r$_7$), the total magnetic moment exhibits a non-monotonic size dependence, and the distribution of the local magnetic moments is highly non-uniform. Moreover, whenever a magnetic ground state is present, multiple magnetic orders lead to very similar total energies, within a range of 25.8~meV (Figure~\ref{fig:mag_mom_en}, insets), which is comparable to thermal energy at room temperature. 

Among the nanoflakes exhibiting a magnetic ground state, local spins are localized on the unpaired Mo $d$-electrons. For r$_4$, the ground state has a total magnetic moment $\mu = 2 \; \mu_B$ (Figure~\ref{fig:r4_mag_mom}). This configuration is ferromagnetic, arising from three local moments aligned parallel to each other. A ferrimagnetic spin pattern  appears in the energetically competing  ($\Delta E < 10~$~meV, see Figure~\ref{fig:mag_mom_en}) configuration with $\mu = 1~\mu_B$, while for $\mu > 1~\mu_B$, only ferromagnetic states appear, where all local magnetic moments are parallel  (Figure~\ref{fig:r4_mag_mom}). The most striking feature is that, in all non-zero magnetic configurations of r$_4$, the local magnetic domains are localized exclusively on three atoms arranged in a triangular pattern. For the configuration with $\mu \geq  2~\mu_B$, these triangularly arranged magnetic domains have nearly the same magnitude. Furthermore, as the magnetization increases, also the remaining Mo atoms in the flake incrementally acquire a non-zero magnetization, as expected.
 
The r$_5$ nanoflake shows a ferrimagnetic ground state with a total magnetic moment of $\mu = 0.5 , \mu_B$, where 4 local spins are aligned parallel while 2 are aligned antiparallel (Figure~\ref{fig:mag_selected}). This configuration is almost degenerate ($\Delta E \approx 12$~meV) to two other ferrimagnetic orderings characterized by  total magnetization of $\mu = 1 \; \mu_B$ and $\mu = 1.5 \; \mu_B$ (Figure~\ref{fig:mag_mom_en}). Similar to r$_4$, the magnetic behavior of the r$_5$ nanoflake is dominated by six Mo edge atoms (two per edge) forming the same pseudo-triangular spatial pattern regardless of the total magnetic moment, with spins aligning either parallel or antiparallel depending on the value of $\mu$ (Figure~\ref{fig:mag_selected} and Figure~S2 in the ESI). Again, for $\mu > 2 \; \mu_B$, the non-zero spins orient parallel to each other, giving rise to ferromagnetic configurations, with the surrounding Mo atoms acquiring a local magnetization increasing with $\mu$.

In the larger nanoflakes, r$_6$ and r$_7$ (Figure~\ref{fig:mag_selected}b,c and Figure~S3, S4 in the ESI), a consistent pattern emerges, similar to r$_4$ and r$_5$, where only a few Mo edge atoms contribute significantly to the total magnetic moment. The ground state of r$_6$ has a total magnetic moment of $\mu = 6 \; \mu_B$. Like r$_4$, this state is ferromagnetic, with all spins aligned parallel to each other (Figure~\ref{fig:mag_selected}b). Other magnetic configurations with $\mu \leq 2.5 \; \mu_B$ are  ferrimagnetic but energetically much less stable, with shifts from the ground state of the order of 300~meV or more. Ferromagnetic states emerge for $\mu \geq 3 \; \mu_B$. Interestingly, the  other ferromagnetic configuration with $\mu = 5.5 \; \mu_B$ is about 100 meV higher in energy compared to the ground state (see Figure~\ref{fig:mag_mom_en}). As shown in Figure~\ref{fig:mag_selected}b, this state features not only fully parallel spin alignment but also total $9$ unpaired spins homogeneously distributed on the nanoflake with similar contribution from each edge.

The ground state of the r$_7$ nanoflake is ferrimagnetic like r$_5$, with a total magnetic moment of $\mu = 2.5 \; \mu_B$, which is less than the r$_6$ flake. In this configuration, 9 Mo edge atoms carry the spins, on the order of 0.5~$\mu_B$ (Figure~\ref{fig:mag_selected}c), arranged in a pseudo-hexagonal pattern along the edges. Similar to r$_5$, all spin along one edge are aligned antiparallel to the other two edges. In this largest nanoflake, the competition between ferrimagnetic and ferromagnetic ordering is significantly higher compared to other structures (Figure~\ref{fig:mag_selected}c and Figure~S4 in the ESI). This is slightly different behavior compared to r$_6$ or smaller sizes where a clear boundary between ferrimagnetic and ferromagnetic ordering occurs. Remarkably,  many states with in this case are the closest in energy to the ground state, spanning a narrow range of 20~meV, which is comparable with thermal energy at room temperature. 

The physical origin of these local magnetic moments is fundamentally linked to the polar discontinuity that emerges at the edge of the nanoflake relative to its bulk-like interior~\cite{Gibertini2014,Gibertini2015,Pino2020}. When an extended crystal terminates at an edge, the local charge balance is disrupted, creating an electrostatically uncompensated net polarization charge. The resulting electrostatic potential gradient drives the accumulation of charge carriers toward the edge to screen polarization charges, ultimately leading to an edge-localized magnetization. This magnetism is strongly size-dependent and only emerges when the edge length is 1.5~nm ($r_4$ nanoflakes) or larger. In these larger systems, we can clearly distinguish between a bulk-like core and an edge region, while below this size, the entire nanoflake is dominated by edge effects and the bulk/edge interface is absent.

The non-monotonic dependence of the total magnetic moment of the nanoflake size is attributed to the energetically competing local spin alignments (ferrimagnetic vs. ferromagnetic ordering) that are nearly degenerate (within a 25~meV range) across different nanoflake sizes. As the structure becomes more extended, the larger edge perimeter provides more Mo atoms near the boundary, leading to a complex landscape where the specific total spin alignment that minimizes the energy and defines the ground state depends on the precise arrangement and number of edge atoms. We expect the magnetization to eventually taper down and disappear once the bulk limit is restored. Based on previous estimates for MoS$_2$ nanoribbons~\cite{Gibertini2015}, the cutoff length is on the order of 50~\AA{}, almost twice the maximum edge size of the triangular nanoflakes explored in this study.

The (pseudo-)triangular spin arrangement is a topological manifestation of the nanoflake geometry, directly correlated with the number of edges and their type. The emergence of local magnetic moments is a consequence of the polar discontinuities at the edges, generating polarization charges $\lambda = \mathbf{P} \cdot \mathbf{\hat{n}}$, where $\mathbf{P}$ is the polarization vector and $\mathbf{\hat{n}}$ is the unit vector orthogonal to the edge. Since each edge is defined by a distinct normal unit vector $\mathbf{\hat{n}}$, it generates a distinct polarization charge density and, consequently, a distinct set of localized, charge-reconstructed edge states on each side. The specific Mo edge atoms hosting the localized magnetic moment are those located at the interface between the inner bulk-like core and the outer edge-like region, where the strong charge reconstruction driven by the H-passivated S-edges is concentrated. Removing a Mo atom from one of these critical corner sites would create S dangling bonds, mirroring the effects seen in non-passivated S-edges, and thus slightly modify the magnetic moments along the affected edge. Likewise, spin localization critically depends on H-passivation, which concentrates the charge reconstruction at these specific Mo atom sites~\cite{Gibertini2014}.

This size-dependent ground-state spin analysis of the \MS nanoflakes reveals a close competition among different magnetic orders. The total magnetic moment of the ground state varies non-monotonically with increasing nanoflake size, and this magnetic behavior is characterized by localized moments confined to specific edge atoms, which form non-trivial spatial arrangements. The presence of nearly degenerate magnetic configurations, with energy differences comparable with thermal energy at room temperature, suggests not only that the ground-state may fluctuate under specific thermodynamic conditions but also indicates a significant potential for tuning the magnetic order using external perturbations. For example, in the most stable ferrimagnetic configuration of r$_4$ flake at $T=0$, an external magnetic field could flip the orientation of the antiparallel spin, enabling a reversible magnetic transition. These possible controllable switching processes highlight the potential of these triangular nanoflakes as active components in spintronic devices, including spin filters and other applications exploiting tunable edge magnetism.

\begin{figure*}
  \centering
  \includegraphics[width=0.95\textwidth]{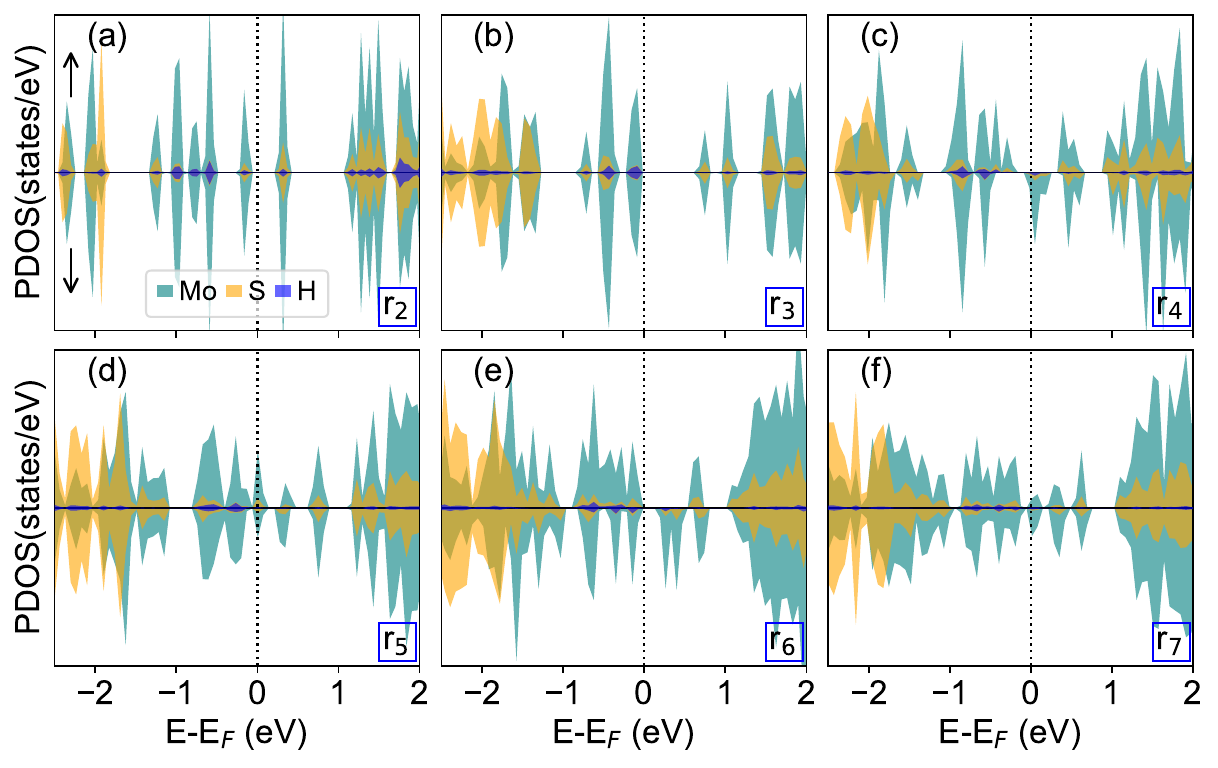}
  \caption{Spin-polarized atom-projected density of states (PDOS) for the magnetic ground state of each \MS nanoflake. The Fermi energy (E$_F$) is set to 0~eV and marked by a vertical dashed bar. }
  \label{fig:dos}
\end{figure*}

The triangular \MS nanoflakes considered in our analysis are often synthesized as colloidal structures in solution. Although this technique has been refined to allow controlled sample preparation~\cite{BERTRAM2006,Huang2000}, irregularities often emerge especially in the edge region. Likewise, H-passivation may be (locally) absent or replaced by other species. Given this variability, we examined the robustness of the effects discussed above -- namely, the presence of magnetic ground states and regular spin patterns formed by specific and spatially ordered edge Mo atoms -- against small structural edge distortions and the absence of H-passivation (see ESI, Figures S5 and S6 for details).

We found a persisting magnetic ordering even in the presence of small deviations from the equilateral nanoflake shape. This is an expected consequence of the polar discontinuity from which this effect originates~\cite{Gibertini2014,Gibertini2015}. This mechanism is electronic, not geometric: the magnetic moment is a response to the uncompensated electrostatic potential, which is primarily dictated by the long-range arrangement of the edge termination and the chemical environment (the S-H bond), and is hardly sensitive to small bond length or angle variations within the edge region.

On the other hand, we noticed that H-passivation plays a critical role: it not only saturates the dangling bonds of S atoms but, importantly, introduces a specific charge environment that dictates the polarization direction, leading to spin localization on the Mo atoms near the edges~\cite{Gibertini2014}. Without H-passivation, the S-edges have a different polarization profile and behave like a closed 1D metallic wire~\cite{Bollinger2001}, suppressing the required charge localization required for the ordered and confined magnetism predicted for their passivated counterparts.

Introducing S vacancies or removing Mo atoms from the non-magnetic core would introduce competing, extrinsic magnetic moments localized at the defect sites. Likewise, substituting Mo with transition metal dopants is expected to introduce strong, localized spins. In all cases, the resulting magnetic ground state would be a complex interplay between the intrinsic edge magnetism (arising from polar discontinuity) and these newly introduced localized moments, likely increasing the total magnetic moment and altering the nearly degenerate spin landscape. Given the complexity and variability of these scenarios, we postpone a dedicated investigation to future work.

\subsection{Electronic properties}
We conclude our analysis by discussing the electronic structure of the energetically most favorable magnetic configuration of all considered \MS nanoflakes in terms of their spin-polarized atom-projected density of states (PDOS, see Figure~\ref{fig:dos}). The smallest nanoflakes, r$_2$ and r$_3$, which are characterized by a non-magnetic ground state, are semiconducting with non-monotonic band gaps of 0.4~eV and 0.8~eV, respectively, consistent with previous works~\cite{Tiutiunnyk2022,Wang2021,BERTEL2019,Wendumu2014}. In line with physical intuition, the band edges are dominated by Mo atoms with small contributions from H (especially in the highest occupied state) and S (at the bottom of the conduction band). S states prevail over Mo states deeper in the valence region, starting from approximately $-2$~eV in both cases. Quantum confinement effects appear in the discrete and energetically separated states characterizing both their occupied and unoccupied regions (Figure~S7). 

Larger \MS nanoflakes, characterized by a magnetic ground state, display a metallic behavior, with Mo $d$-states crossing the Fermi energy (Figure~\ref{fig:dos}). Interestingly, the even-numbered nanoflakes (r$_4$ and r$_6$) exhibit a significantly lower density of states at the Fermi level compared to their odd-numbered counterparts. In particular, r$_6$ exhibits a small gap of a few tens of meV in each spin channel. In r$_5$ and r$_7$, where the metallicity is more pronounced, an even larger gap of the order of 100~meV is visible in the spin-down channel, as expected in the presence of a ferromagnetic ground state. The larger number of Mo and S atoms compared to the passivating hydrogens is reflected in the atom-resolved contributions to the PDOS, which are overwhelmingly dominated by the metal and chalcogen species (Figure~\ref{fig:dos}). Nonetheless, similar to the smaller structures, even in the larger nanoflakes, the gap region hosts predominantly Mo contributions while S states prevail deeper in the valence region, here starting from approximately 1.5~eV, and higher in the conduction band, starting from 1~eV, where they are hybridized by Mo orbitals.

This analysis on the electronic structure of the considered \MS nanoflakes consistently reflects the previous discussion on their magnetic properties. The semiconducting nature of the smallest, non-magnetic nanostructures (r$_2$ and r$_3$) is governed by quantum confinement. Conversely, the metallic behavior predicted for the larger magnetic nanoflakes (r$_4-$r$_7$) results from the unpaired spins in the Mo $d$-orbitals crossing the Fermi level. The larger magnetization and the near-degeneracy of magnetic orders of these larger systems are directly manifested in the complex spin-split PDOS near the Fermi level, supporting their potential as platforms for tunable edge magnetism. We note in passing that the results displayed in Figure~\ref{fig:dos} are obtained at the PBE+SP+D3 level of theory, which excludes static and dynamic correlations that are crucial for a quantitative description of magnetic nanostructures. As such, our results are intended solely to provide a \textit{qualitative} picture of the trends arising in the PDOS of the considered nanoflakes, without claiming quantitative predictive accuracy.

\section{SUMMARY AND CONCLUSIONS}
In summary, we systematically investigated the intrinsic magnetic properties of triangular \MS nanoflakes using spin-polarized DFT. By varying the edge length of S-terminated nanoflakes with hydrogen passivation, we identified a critical size threshold of approximately 1.5~nm, below which the structures remain non-magnetic. Larger nanoflakes exhibit multiple, closely competing magnetic states arising from localized moments on specific Mo edge atoms. All magnetic configurations are metallic, with unpaired Mo $d$-orbitals crossing the Fermi level. The robustness of this edge-induced magnetism is underscored by finding that edge-localized magnetic moments persist even in structurally modified, non-equilateral geometries. In contrast, removing hydrogen passivation leads to a disordered spin landscape, where magnetic moments become randomly distributed across both Mo and S edge atoms. 

In conclusion, our results demonstrate that S-terminated and H-passivated triangular \MS nanoflakes represent a stable, experimentally accessible, and promising platform for nanoscale spin control. The robust and localized magnetic moments in these nanoflakes can be manipulated by external electric or magnetic field to realize advanced spintronic applications, primarily by using electrostatic gating. For instance, if  magnetic field with alternating component is applied, it will induce spin rotation and hence results in spin manipulation for the design of low-dimensional spintronic devices such as spin filters and memory elements. Similarly alternating electric field allows manipulation of the spin state at will via electric-dipole-induced spin resonance and therefore can be explored as spin qubit. Given the analogies among TMDs, we anticipate that the physics predicted here also exists in other members of this family, such as W-based nanoflakes. However, a detailed analysis of different metallic and/or chalcogen species, as well as the influence of defects Mo and S vacancy and functionalization, requires dedicated analysis, which will be pursued in future work.

%%%%%%%%%%%%%%%%%%%%% End matter %%%%%%%%%%%%%%%%%%%%% 
\section*{Author contributions}
\textbf{Surender Kumar}: conceptualization, methodology, calculations and formal analysis, writing original draft; \textbf{Stefan Velja}: software,  review \&  editing; \textbf{Muhammad Sufyan Ramzan}:  review \&  editing;
\textbf{Caterina Cocchi}: conceptualization, resources, supervision, writing – review \& editing.

\section*{Conflicts of interest}
There are no conflicts of interest to declare.

\section*{Data Availability}
The raw data presented in this study are available free of charge in Zenodo \href{https://doi.org/10.5281/zenodo.17457576}{DOI: 10.5281/zenodo.17457576} [record: 17457576].
%\section*{Acknowledgment}

%%%%%%%%%%%%%%%%%%%%% biblography %%%%%%%%%%%%%%%%%%%%% 
\bibliography{main}

\end{document}